\documentclass[useAMS,usenatbib]{mn2e}
\usepackage{amssymb,amsmath,epsfig,times, natbib,color,threeparttable}
\pdfoutput=1

\title[NGC1600 Bondi Radius]{Probing within the Bondi radius of the ultramassive black hole in NGC 1600}
\author[J. Runge] {\parbox[]{6.5in}{
      J. Runge$^1$\thanks{Email: 
    jmr0062@uah.edu}, {S. A. Walker$^1$}}\\
     \footnotesize
     $^1$Department of Physics and Astronomy, The University of Alabama in Huntsville, 301 Sparkman Drive NW, Huntsville, AL 35899, USA \\
}

\date{}

\begin{document}

\maketitle

\begin{abstract}
We present deep (250 ks) \textit{Chandra} observations of the nearby galaxy group NGC 1600, which has at its centre an ultramassive black hole (17$\pm$1.5 billion $\textup{M}_{\odot}$). The exceptionally large mass of the black hole coupled with its low redshift makes it one of only a handful of black holes for which spatially resolved temperature and density profiles can be obtained within the Bondi radius with the high spatial resolution of \textit{Chandra}. We analyzed the hot gas properties within the Bondi accretion radius R$_\textup{B}=1\farcs2 - 1\farcs7= 0.38 - 0.54~\textup{kpc}$. Within a $\sim\!3$~kpc radius, we find two temperature components with statistical significance. Both the single temperature and two temperature models show only a very slight rise in temperature towards the centre, and are consistent with being flat. This is in contrast with the expectation from Bondi accretion for a temperature profile which increases towards the centre, and appears to indicate that the dynamics of the gas are not being determined by the central black hole. The density profile follows a relatively shallow $\rho\propto~r^{-[0.61\pm0.13]}$ relationship within the Bondi radius, which suggests that the true accretion rate on to the black hole may be lower than the classical Bondi accretion rate. 
\end{abstract}

\section{Introduction}

\cite{Thomas2016} have recently reported the discovery of a 17$\pm$1.5 billion solar mass ultramassive black hole (UMBH) at the centre of the group galaxy NGC 1600. The black hole’s colossal mass, and relatively close proximity (z=0.0156) mean that it is one of only a handful of black holes whose Bondi radius is large enough to be spatially resolved by \textit{Chandra} (R\textsubscript{Bondi}=0.54 kpc=1.7 arcsec), and the only ultramassive (M$>$10 billion $\textup{M}_{\odot}$) black hole for which this is the case. Inside the Bondi radius (R$_\textup{B}$ = 2GM$_{BH}$ /c$^2_s$ ) the gravitational potential of the black hole of mass M$_{BH}$ is dominant over the thermal energy of the surrounding gas (whose sound speed is c$_s$), and the gas inside this radius is expected to be accreted by the black hole.

At present only three other black holes have been probed within the Bondi radius with \textit{Chandra}: Sgr A$^*$ at the centre of our galaxy (\citealt{Wang2013}), and the central black holes in the galaxies NGC 3115 (\citealt{Wong2011,Wong2014}: hereafter W14) and M87 (\citealt{Russell2015,Russell2018}: hereafter R15 and R18 respectively), with each exhibiting different behaviour. Such observations are extremely challenging, requiring the central black hole to be massive, nearby, and its X-ray emission to not be piled up.
\begin{figure*}
	\includegraphics[width=0.49\textwidth]{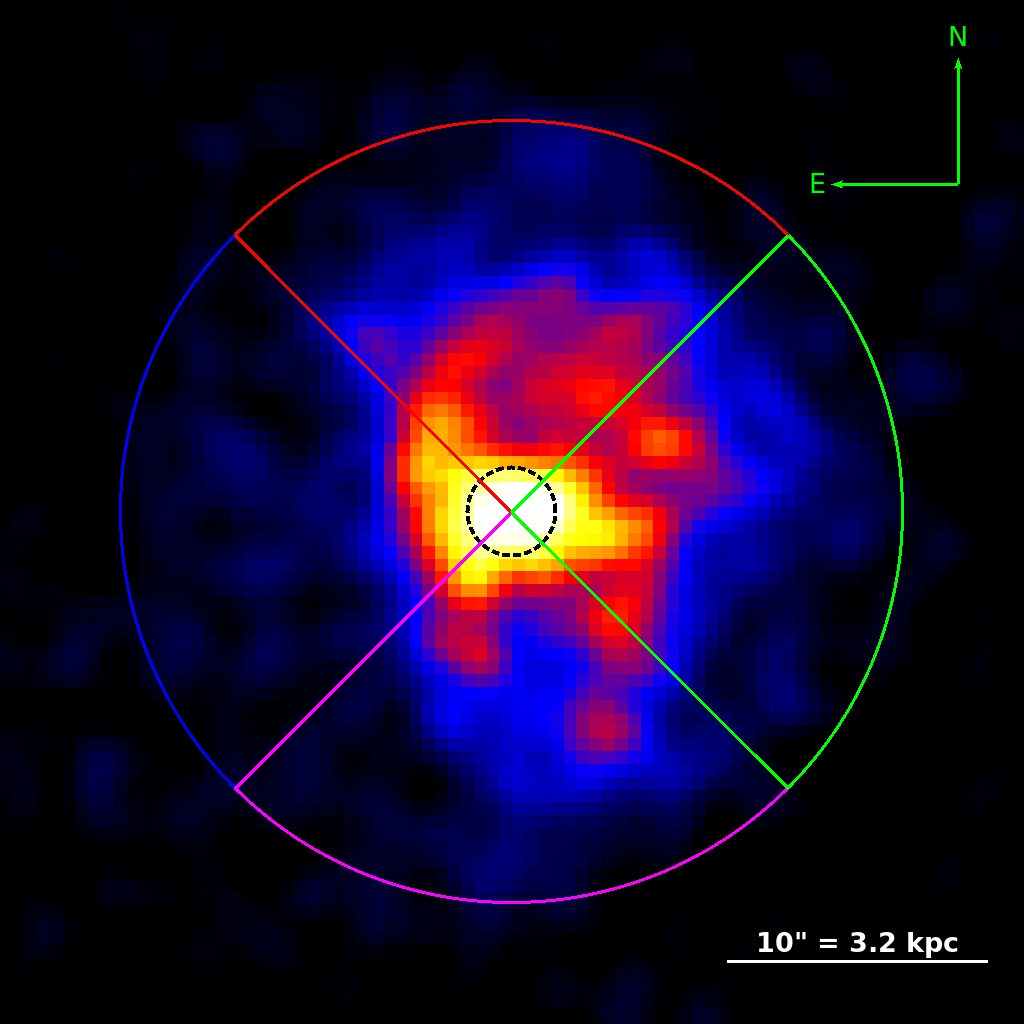}
	\includegraphics[width=0.49\textwidth]{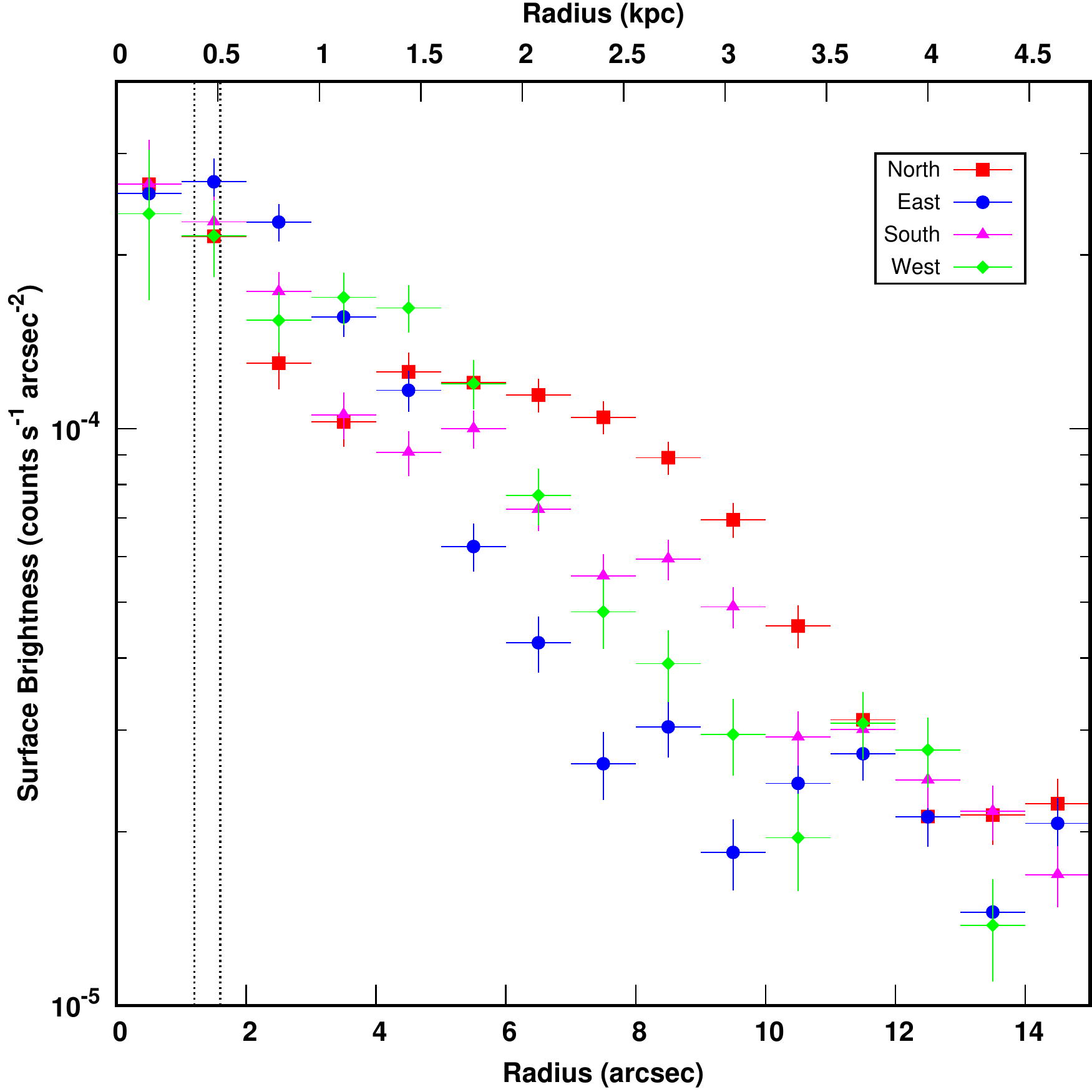}
	\caption{Left: Smoothed soft band (0.5 - 1.2 keV) \textit{Chandra} image of NGC 1600 in the native pixel scale of 0\farcs49. The black-dashed region shows the Bondi radius at 1.7\arcsec (0.54 kpc). Regions showing the quadrants used to analyze the surface brightness profile of NGC 1600. Right: Surface brightness profile for the North, East, South, and West quadrants of NGC 1600. Colors are the same as the regions shown on left. The vertical dashed lines indicate the range of the Bondi radius.}
	\label{fig:subimage}
\end{figure*}

NGC 1600 has a very similar environment to M87: both lie at the centres of galaxy groups with total masses around $1.5\times10^{14}$ M$_{\odot}$ (\citealt{Walker2013}, \citealt{Thomas2016}). However NGC 1600 has several advantageous characteristics. Firstly, the central AGN emission is extremely faint; a study of the orginal 45ks Chandra observation by \citet{Sivakoff2004} did not find any point source associated with the central black hole, and in this work we find an upper limit of $L_{X,AGN} < 6\times10^{37}$ erg s$^{-1}$ in the 0.5-6.0 keV band. This avoids the complication found for M87, whose central AGN is very bright and is piled up in many observations taken with Chandra (\citealt{Russell2015,Russell2018}), making it difficult to measure temperatures and densities inside its Bondi radius. The orbit superposition method used in \cite{Thomas2016} provides a very precise black hole mass value (17$\pm$1.5 billion $\textup{M}_{\odot}$) for NGC 1600. This is far more precise than the black hole mass for M87, where the estimates range from 3.5-6.6 billion $\textup{M}_{\odot}$ (\citealt{Walsh2013,Gebhardt2011,EHT19}), leading to a factor of 2 uncertainty in the Bondi radius of M87.

NGC 1600 holds several advantages over NGC 3115 as well. Similar to M87, the Bondi radius value is highly uncertain for NGC 3115 (where \citealt{Kormendy96} and \citealt{Emsellem99} only roughly constrain the black hole mass to be 1-2 billion $\textup{M}_{\odot}$). The much greater mass of NGC 1600 (which lies in the centre of a galaxy group) compared to NGC 3115 (which is a field lenticular galaxy) means that the X-ray luminosity of the gas in NGC 1600 is much higher (by a factor of $\approx\!40$). Therefore the fraction of contaminating emission from low mass X-ray binaries (LMXBs) is much smaller (like it is for M87) and can be easily removed through spectral modelling. 


The low AGN luminosity gives the black hole an extremely low Eddington fraction (L$_X$/L$_{Edd} <10^{-10}$), similar to NGC 3115. Were the UMBH to be powered by Bondi accretion, with a standard radiative efficiency of 10 percent, the resulting luminosity would be $\sim10^{45}$ erg s$^{-1}$ (i.e. the level of a quasar), over eight orders of magnitude higher than is observed. This means that the accretion onto the black hole is extremely inefficient. We would therefore expect the black hole to be in a hot, radiatively inefficient accretion flow (RIAF) mode, instead of a thin accretion disk, with the gas temperature rising inwards ($\textup{T}\propto r^{-1}$, \citealt{NaraMc08}). The low efficiency is consistent with an advection dominated accretion flow (ADAF), where the advection of energy dominates as the gas cannot radiate energy efficiently due to its low density. Alternate models predict that the gas is prevented from reaching the black hole and instead is blown out of its sphere of influence through winds (ADIOS, e.g. \citealt{BB04}), while others predict the gas to form convective eddies (CDAF, e.g. \citealt{Narayan00}), again reducing the accretion rate. Our new observations and subsequent analysis hope to reveal some insights into the accretion mechanics at work within NGC 1600.
\begin{table}
    \centering
     \caption{Overview of the \textit{Chandra} observations used in this study.}
    \begin{threeparttable}
        \begin{tabular}{l|r|c}
     \hline
     \hline
     Obs. ID & Date & Exposure (ks)  \\
      \hline
       4283  & 9/18/02  & 28.5  \\
        4371  & 9/20/02  & 28.5   \\
        21374 & 12/3/18  & 26     \\
        21375 & 11/28/19 & 42     \\
        21998 & 12/3/18  & 14    \\
        22878 & 11/25/19 & 46     \\
        22911 & 11/1/19  & 31    \\
        22912 & 11/2/19  & 36     \\
        Stack &          & 252  \\
      \hline
      \end{tabular}
     \label{tab:obs}
     \end{threeparttable}
        
    \label{table:observations}
\end{table}

In this paper, we use 252 ks of \textit{Chandra} X-ray observations of NGC 1600 in order to determine the temperature and density profiles within the Bondi radius of the central black hole. Section 2 details the observations, data reduction, and spectral analysis procedure. In Section 3, we present the radial profiles for temperature, density, entropy, and the cooling time within the Bondi radius. Section 4 discusses the implications of our findings and how they compare to similar studies done at this scale. Throughout the paper, we assume a $\Delta$CDM cosmology with $\Omega_\Delta=0.7$, $\Omega_m=0.3$, and $H_0=70$~km s$^{-1}$ Mpc$^{-1}$. At the redshift of NGC 1600, 1 arcsecond corresponds to 0.32 kpc. All errors are 1-$\sigma$ unless otherwise noted.

\section{Data}

\subsection{Data Reduction}

For this study of NGC 1600 (Figure~\ref{fig:subimage}), we use a total of eight \textit{Chandra} observations (Table~\ref{table:observations}): Six of our own observations totalling 195ks taken between 2018 and 2019 (ObsIDs 21374, 21375, 21998, 22878, 22911, and 22912; PI Walker) and two archival observations from 2002 (ObsIDs 4283 and 4371; PI Sarazin) for a total of 252 ks. All data were reprocessed using the \texttt{chandra\_repro} script of \texttt{CIAO 4.12} and \texttt{CALDB 4.9.1}. Background light curves were analyzed to filter out any time periods affected by flares.
In order to properly analyze the diffuse X-ray emission, point sources needed to be removed. These sources were detected using \texttt{CIAO wavdetect} on the merged broad energy image and excluded from the analysis. There was no detection of a central point source indicative of AGN activity. To put a limit on the X-ray luminosity of the central AGN, we follow the method of W14 of fitting the surface brightness profile shape while including a point source component in the surface brightness model. We measure an X-ray luminosity of the central AGN of $L_{X,AGN} = 6\times10^{37}$ erg s$^{-1}$ in the 0.5-6.0 keV band.

\subsection{Surface Brightness Profile}
NGC 1600 has an asymmetrical morphology; extending further North and South than East or West. Examining the surface brightness profile in these four quadrants (Figure~\ref{fig:subimage}), we see a number of things. Firstly, there is little scatter among the profiles within 3\arcsec (1 kpc), indicating that any AGN inflated cavities are not significantly affecting the distribution of the ICM on these scales. Secondly, there is greater variability between 3\arcsec and 12\arcsec (1-4 kpc) where we see what appear to be the inner rims of cavities to the north and south. \citet{Russell2015} find similar behavior for M87 when dividing their analysis into sectors, with the radial scatter in the surface brightness being relatively small in the central 0.3 kpc, and so like \citet{Russell2015} we proceed with azimuthally averaged temperature and density profiles.

\subsection{Spectral Analysis}

We extracted spectra in circular annuli centered on the optical peak from archival \textit{HST} images (ra: 67.9161163 dec: -5.0862464). It is separated from the X-ray emission peak by $0\farcs25$, and we find no difference in our results if we centre our analysis on the X-ray peak instead. The spectra were analyzed using the X-ray Spectral Fitting Package (\texttt{XSPEC}).

\begin{figure}
	\includegraphics[width=\linewidth]{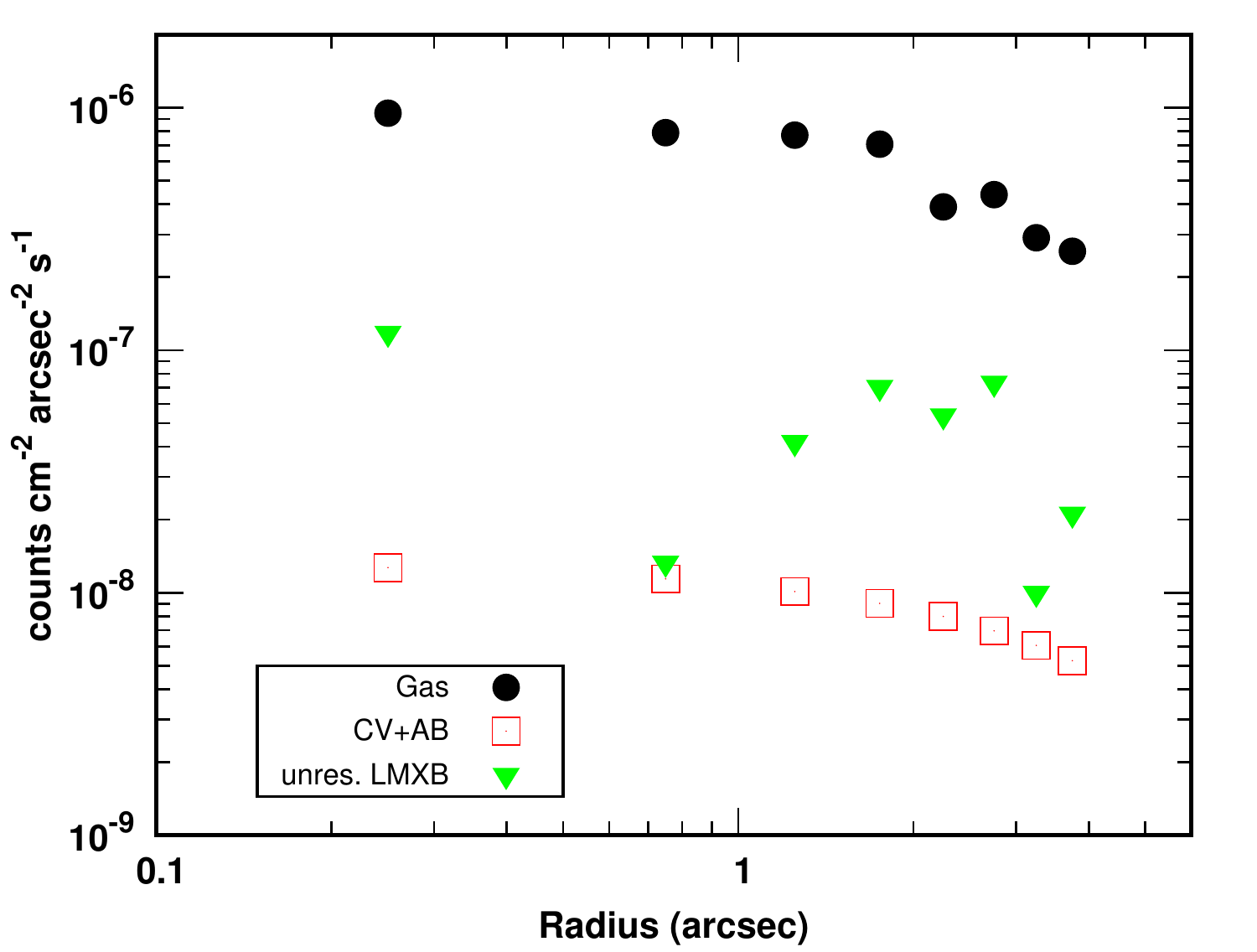}
	\caption{Surface brightness profiles for the hot gas (black circles), CV/ABs (red, open squares), and LMXB (green triangles) components of NGC 1600 in the inner 4\arcsec.}
	\label{fig:X_comps}
\end{figure}

As outlined in W14, the X-ray emission has three main contributions: the diffuse hot gas component, unresolved low-mass X-ray binaries (LMXBs), and stellar emission from cataclysmic variables and coronally active binaries (CV/ABs). We follow a similar process as outlined by W14 in order to estimate the CV/AB contribution. Taking the K-band luminosity found by 2MASS, and assuming that the K-band surface brightness profile shape follows that from an archival \textit{HST} WFC3 F475X image, we estimated the K-band surface brightness profile. Using the $L_\textup{X} - L_\textup{K}$ scaling relation derived for M32 (\citealt{Rev07}), the CV/AB normalizations for 0\farcs5 annuli were determined. Owing to the much higher mass and X-ray surface brightness of NGC 1600 compared to NGC 3115, we find that (as is also the case with M87), such a component is negligible ($<\!2\% $) when compared to the gas component (Figure~\ref{fig:X_comps}), and that including such a background component in the spectral modelling has no effect on our spectral fits. Therefore, we only consider contributions from the hot gas and LMXBs in this paper.

For each observation, spectra were extracted from 0\farcs5 annuli using \texttt{CIAO specextract}, which creates the source and background spectra along with the associated ARF and RMF response files. Background was extracted locally from a region far away from the center of NGC 1600 where the surface brightness profile was flat. We then merged with the spectra from the other observations using \texttt{CIAO combine\_spectra}, which sums the multiple spectra and combines the background spectra as well as the response files. The merged spectra in the 0.5-7.0 keV range were fitted with a two component absorbed (\texttt{PHABS}) model: a thermal (\texttt{APEC}) model for the gas, and a power law (\texttt{POWERLAW}) with a fixed slope of 1.6 for the LMXBs (consistent with the value of 1.56$^{+0.03}_{-0.03}$ we obtain from the combined spectrum of all of the point sources we identify in NGC 1600). As shown in Fig. \ref{fig:X_comps}, the contribution to the total emission from LMXBs in small. Varying the powerlaw index through the range 1.3-1.9 in our spectral fits has no affect on our results. The Galactic column density was fixed at a value of $N_{\textup{H}}=3.14\times10^{20}$ cm$^{-2}$ (\citealt{DL90}). We used the \texttt{wilm} abundance table (\citealt{Wilms2000}). All spectra were fitted using the modified C-statistic (\citealt{Cash1979}).


\section{Results}
\begin{figure}
	\includegraphics[width=\linewidth]{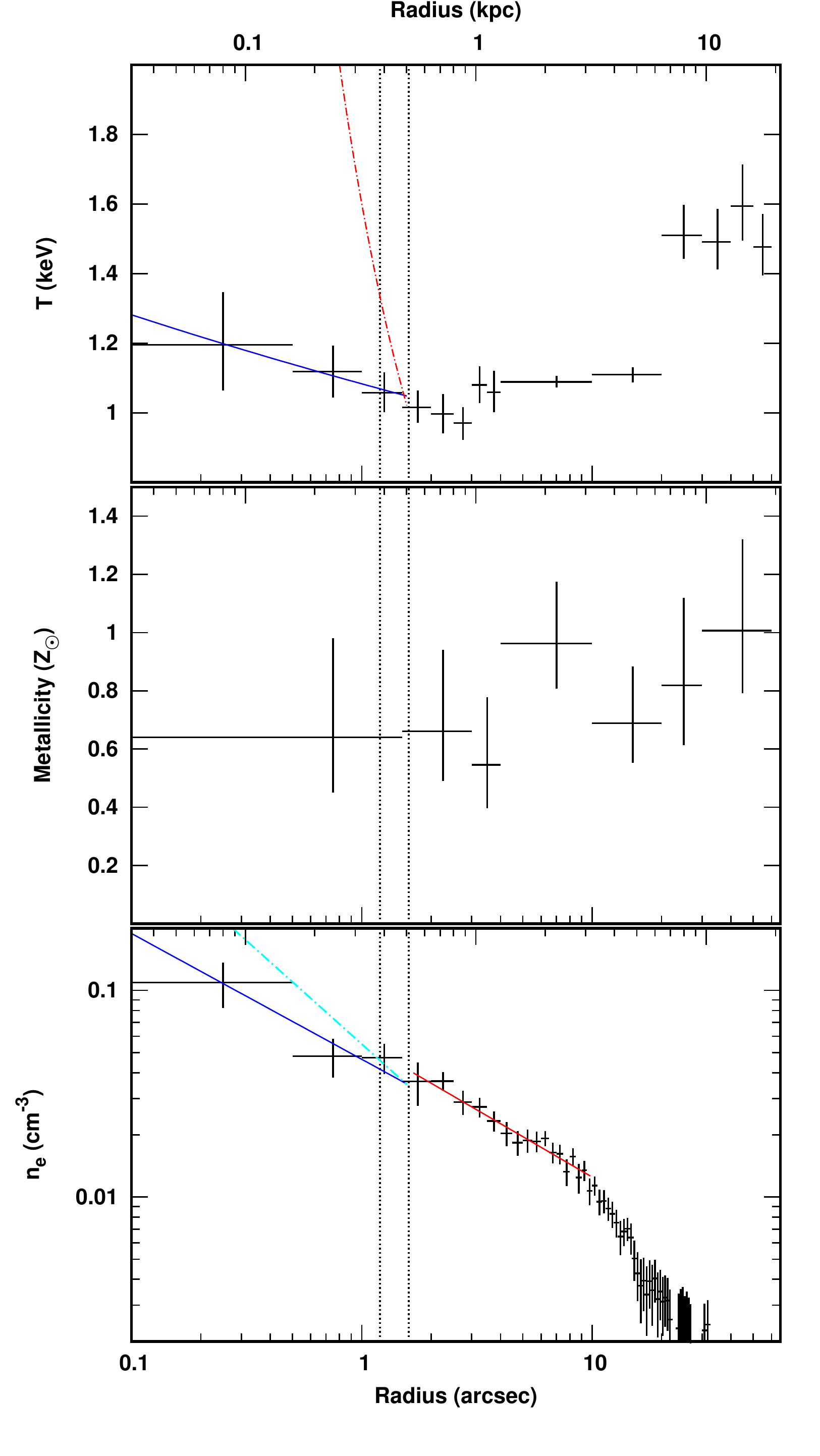}
	\caption{Top: Temperature profile using a single temperature model using 0\farcs5 bins. The Bondi radius region (1\farcs2 - 1\farcs7) is marked by dashed lines. The blue line is the best fit powerlaw within the Bondi radius and has a power law index of 0.07; for comparison, an index of 1 is shown as a red, dot-dashed line. Middle: Metallicity profile.  Bottom: Deprojected density profile using 0\farcs5 bins. The best fit powerlaw within the Bondi radius is shown as the blue line and has a power law index of 0.61; a cyan dot-dashed line with power law index of 1 is shown for comparison. The red line shows the best fit powerlaw outside the Bondi radius out to a radius of 10 arcseconds, and has a power law index of 0.65.}
	\label{fig:multiplot}
\end{figure}

\subsection{Temperature Profile}
Figure~\ref{fig:multiplot} (top panel) shows the projected temperature profile using 0.5\arcsec wide radial bins for the inner 4\arcsec (1.28 kpc) and 6-10\arcsec wide bins out to 60\arcsec (19.2 kpc). Within the Bondi radius, the temperature profile demonstrates a mild increase towards the core, consistent with a power law of $\textup{T} \propto r^{-0.073\pm0.011}$ (shown by the solid blue line in the top panel of Fig. \ref{fig:multiplot}).

In W14's study of NGC 3115 it is found that the hotter gas component within the Bondi radius gives a power law relation of $\textup{T} \propto r^{-0.44^{+0.29}_{-0.33}}$. This value is steeper than what we find, but both are still shallower than the $\textup{T}\propto r^{-1}$ relation expected for classic Bondi accretion (shown as the red dot dashed line for comparison in the top panel of Figure~\ref{fig:multiplot}) or the lower limit of the powerlaw index of  0.6 found for Sgr A$^*$ (\citealt{Wang2013}). 

If metallicity is set to a free parameter, we find that it stays relatively constant (Figure~\ref{fig:multiplot}, middle panel). While it is possible that there is Fe bias due to using a single temperature model (\citealt{Buote00}), it does not result in extremely, sub-solar metallicity abundances. 


\subsubsection{Two Temperature Model}
\begin{figure}
	\includegraphics[width=\linewidth]{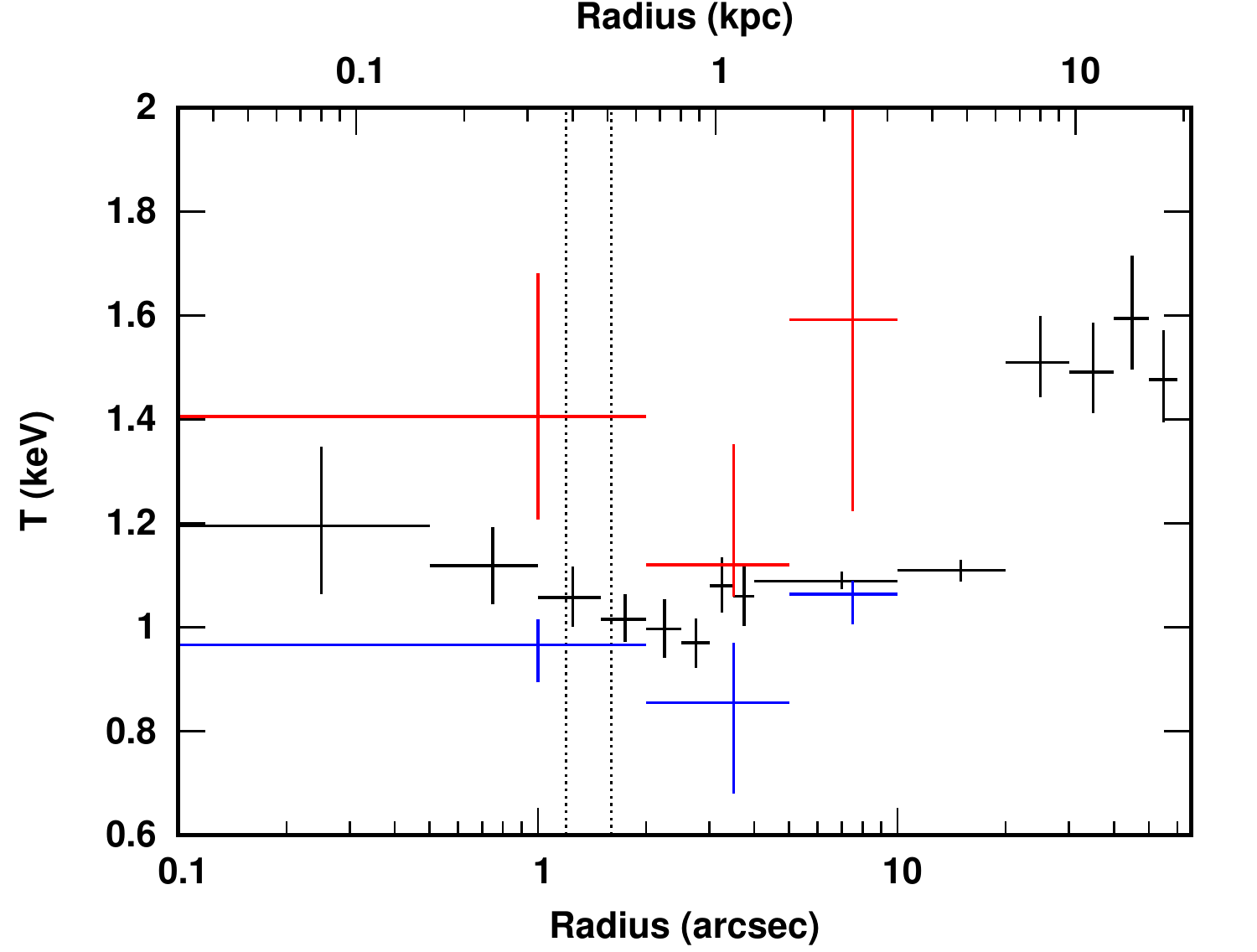}
	\caption{Temperature profile with hot (red) and cold (blue) gas components of two temperature model over-plotted on single temperature profile.}
	\label{fig:2temp}
\end{figure}

In order to check for a possible multiphase system, we fit a two-temperature model to annuli within the inner 10\arcsec. Larger annuli were used than for the single temperature fits to achieve measurements with sufficient statistical accuracy. 
The metallicity was frozen to its best fit value from the single temperature fits while the temperature and normalization for the two \texttt{APEC} models were allowed to vary. For the inner $\sim\!2$kpc, an F-test reveals a $<0.04\%$ probability that this secondary component is detected by chance. This probability increases to $\sim\!8\%$ for the outermost annuli (5 - 10\arcsec). Figure~\ref{fig:2temp} shows the hot and cold gas components from the two-temperature model. The temperature of both components still rises slightly towards the center as previously seen in the single temperature profile, though the significance of this increase is limited by the large error bars.

\begin{table}
    \centering
    \caption{Comparison of azimuthally averaged deprojected electron density profile power law indices among objects with sub-Bondi radius observations. The values are for single temperature fits to the data. NGC 3115 values are from \citet{Wang2013} while M87 values are from \citet{Russell2015}. }
    \begin{threeparttable}
        \begin{tabular}{l|c|c}
     \hline
     \hline
     Object & $\beta$ (r $<$ $r_B$) & $\beta$ (r $>$ $r_B$) \\
      \hline
           \rule{0pt}{2ex}  NGC 3115 & $0.62^{+0.26}_{-0.38}$ & $1.34^{+0.20}_{-0.25}$ \\
           \rule{0pt}{3ex} M87 & $1.0^{+0.20}_{-0.20}$ & $0.41^{+0.04}_{-0.04}$  \\
           \rule{0pt}{3ex} NGC 1600 & $0.61^{+0.13}_{-0.13}$ & $0.65^{+0.04}_{-0.04}$ \\
      \hline
      \end{tabular}
     \label{tab:my_label}
     \end{threeparttable}
    \label{table:indices}
\end{table}
\subsection{Deprojected Density Profile}

Using circular radial bins of 0\farcs5 out to 60\arcsec and excluding point sources, we were able to obtain the deprojected density profile using the DSDEPROJ deprojection routine method (\citealt{Russell2008}).
The deprojected electron density profile for NGC 1600 is shown in the bottom panel of Figure~\ref{fig:multiplot}. Fitting a power law to the density profile within the Bondi radius ($r < 1\farcs7$) gives $\rho\propto~r^{-[0.61\pm0.13]}$, shown by the solid blue line in Fig. \ref{fig:multiplot}, bottom panel. 

In Table~\ref{table:indices} we compare the density profile powerlaw indices for fitting inside and outside the Bondi radius with the equivalent values obtained for NGC3115 in W14 and for M87 in R15. We compare to R15 rather than R18 as the former presents an azimuthally averaged analysis which more closely matches the analysis we perform, as we lack the data quality needed to perform the azimuthally resolved analysis in sectors that is presented in R18. We find that the power law index for the density for NGC 1600 inside the Bondi radius is consistent with that which W14 finds for their single-temperature model for NGC 3115 but shallower than the R15 estimate for M87 (see Table~\ref{table:indices} for a comparison). 

If we look at the density profile outside the Bondi radius ($1\farcs7 < r < 20\arcsec$), the power law index is consistent with being the same as inside the Bondi radius,  $0.65\pm0.04$. This contrasts with M87, where the density profile is flatter outside the Bondi radius, and with NGC 3115, where the density profile is steeper outside the Bondi radius. As is discussed in R15, the X-ray emission around the M87 system is highly disturbed due to multiple generations of cavities created by AGN feedback, which has more greatly affected the shape of the surface brightness profile compared to NGC1600 (which has only one pair of cavities). This greater amount of AGN feedback in M87 appears to have pushed more gas outwards, thus causing the density gradient to be flatter outside the Bondi radius. NGC 3115, on the other hand, shows little to no AGN feedback activity, resulting in a steeper density gradient outside the Bondi radius. While NGC 1600 does show some evidence of AGN feedback (i.e. possible cavities), it is nowhere to the extent of M87; therefore it is not unsurprising that we observe a density profile outside the Bondi radius with a power law index between the values found for M87 and NGC 3115. 

\subsection{Entropy}
\begin{figure}
	\includegraphics[width=\linewidth]{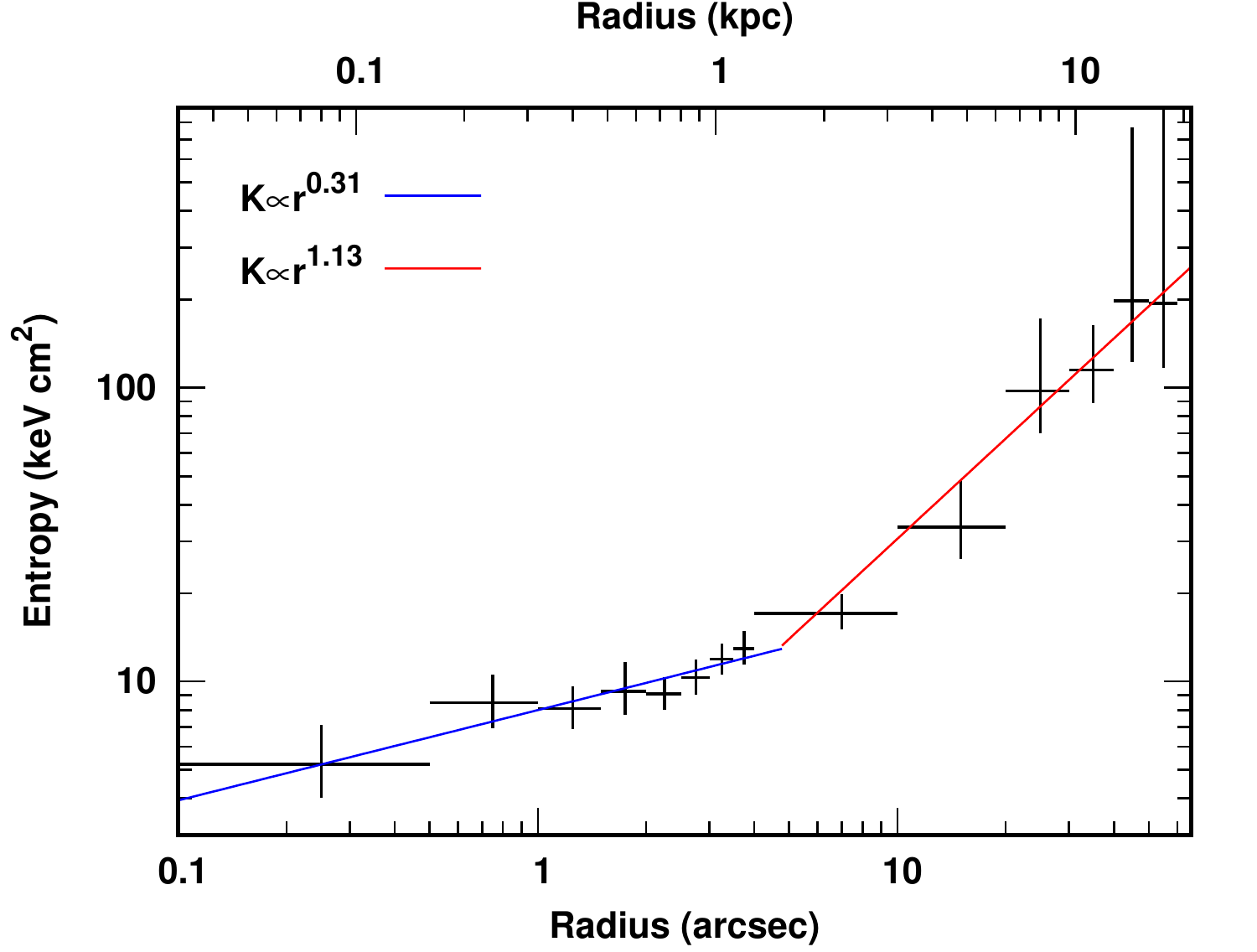}
	\caption{Entropy profile for NGC 1600 fitted with a broken power law. The power law indices are 0.31 (blue line) for the inner portion and 1.13 (red line) for the outer portion.}
	\label{fig:Entropy}
\end{figure}

It has been shown that thermal instabilities are prevalent in systems whose central entropy is $<\!30~\textup{keV}~\textup{cm}^2$ (\citealt{Cavagnolo08,Rafferty08,Voit08}). We find the entropy profile ($\textup{K}=\textup{kT}/\textup{n}_e^{2/3}$) using the temperature and deprojected density profiles (Fig.~\ref{fig:Entropy}). The entropy is below the $30~\textup{keV}~\textup{cm}^2$ threshold out to $\sim\!3$~kpc. Outside 1kpc, the profile follows a $\textup{K}\propto r^{1.13\pm0.21}$ relation, which is in agreement with $\textup{K} \propto r^{0.99\pm0.11}$ that \cite{Babyk2018} found for a sample of 40 early type galaxies and 110 galaxy clusters. Like previous studies have shown (\citealt{Donahue06,Cavagnolo09,Voit16}), the entropy profile flattens in the central region, and can be fit with a broken powerlaw. For $r<1$~kpc, the entropy profile becomes $\textup{K}\propto r^{0.31\pm0.05}$; shallower than the average relationship of $\textup{K}\propto~r^{2/3}$ found for samples of systems (\citealt{Panagoulia14,Babyk2018}), but within the scatter of the sample of systems explored in \citet{Babyk2018}. It should be noted that we are working on a smaller scale compared to that of \cite{Babyk2018} and \cite{Panagoulia14}, who work on scales from $\sim$1 to 1000 kpc, with a majority of data points above 10 kpc. For instance in \citet{Babyk2018} the central slope of $\textup{K}\propto~r^{0.67}$ is found by fitting a powerlaw to a much larger radial range, namely the central $\approx$15 kpc. If we fit a powerlaw to the same radial range as \citet{Babyk2018} (the central 15 kpc) we do indeed obtain a powerlaw index consistent with  $r^{0.67}$.
\subsection{Cooling time and dynamical time}
\begin{figure}
	\includegraphics[width=\linewidth]{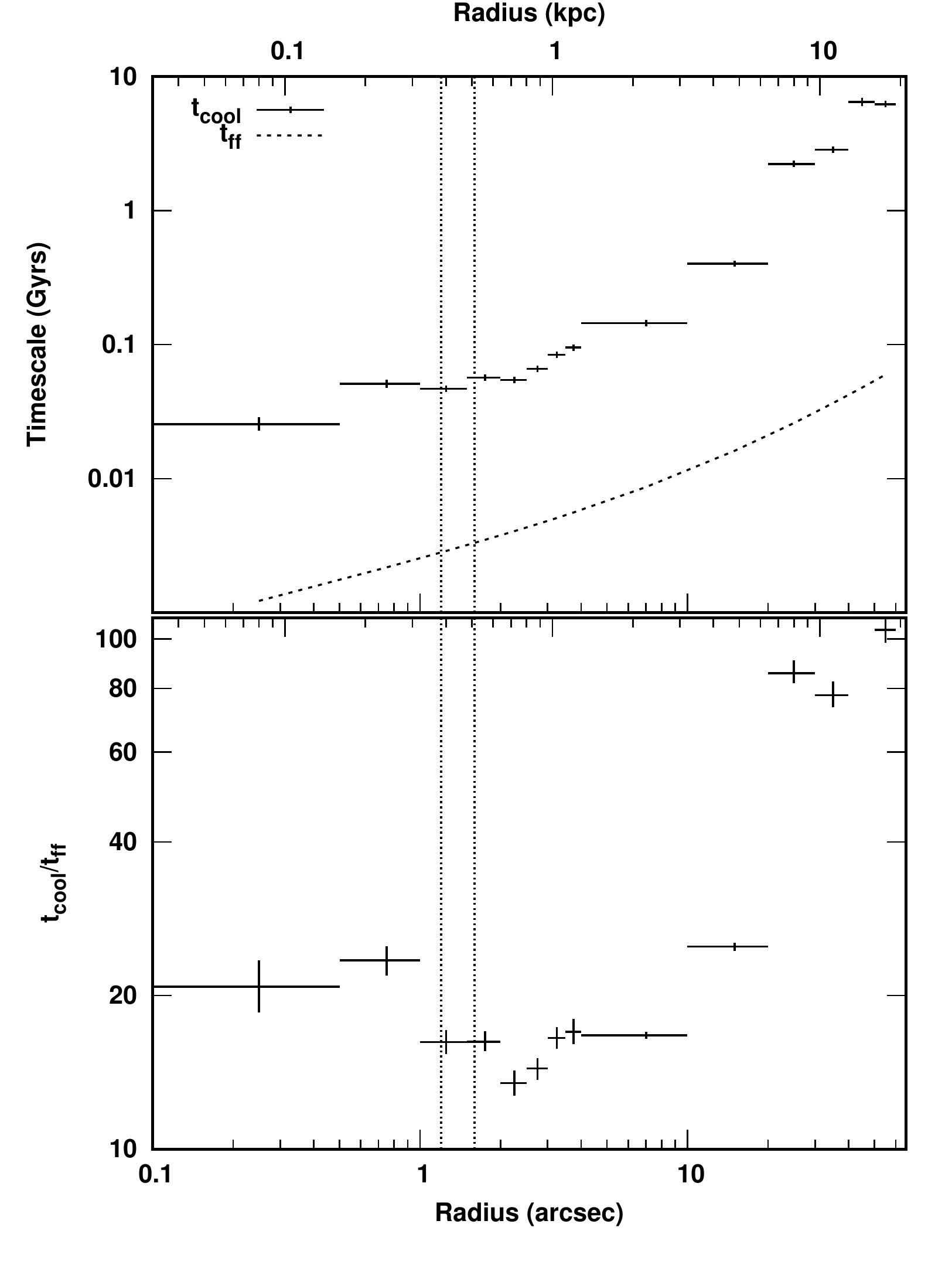}
	\caption{Top: radiative cooling time profile and dynamical free-fall time profile. Bottom: ratio of cooling time to free-fall time.}
	\label{fig:time}
\end{figure}

 Multi-temperature structure may also appear when the cooling time becomes sufficiently short with respect to the dynamical free-fall time. Therefore, we have calculated the cooling time and the free-fall time. The radiative cooling time, $t_{\textup{cool}}=(3/2)\textup{nkT/n}^2\Lambda$, was calculated from the temperature and density profiles. The free-fall time was calculated using a best fit mass profile. We assume the mass density follows a Navarro-Frenk-White (NFW) profile (\citealt{NFW97,Schmidt07})
\begin{equation}
\rho(r) = \frac{\rho_0}{r/r_{\textup{s}}(1+(r/r_{\textup{s}}))^2}
\end{equation}
\begin{equation}
\rho_0 = 200\rho_cc^3_{200}/3(ln(1+c_{200})-c_{200}/(1+c_{200}))
\end{equation}
where $\rho_c$ is the critical density of the universe, and $\textup{c}_{200} =\textup{r}_{200} /\textup{r}_s$. Temperatures are calculated in each annulus with the deprojected density profile and NFW mass profile. By comparing these calculated temperatures to the measured temperatures, we find the best fit NFW mass profile which has the lowest $\chi^2$ statistic, defined as
\begin{equation}
\chi^2=\sum_{i}\frac{(T_{calculated,i}-T_{actual,i})^2}{\sigma^2_{T_{actual,i}}}
\end{equation}
Figure~\ref{fig:time} shows the cooling time, free-fall time, and the $t_{\textup{cool}}/t_{\textup{ff}}$ profile calculated for NGC 1600. The ratio reaches a minimum of $t_{\textup{cool}}/t_{\textup{ff}}=13.48\pm0.77$ at 0.72 kpc. Akin to what \citet{Russell2015} observed for M87, we see an increase of $t_{\textup{cool}}/t_{\textup{ff}}$ in the central region.
\subsection{Bondi radius and mass accretion rate}
The radius at which the SMBH's gravitational potential becomes dominant is determined by its central gas temperature (\citealt{Bondi52}). Assuming spherical symmetry, the Bondi radius is given by the following:
\begin{equation}
\frac{\textup{R}_{\textup{B}}}{\textup{kpc}}= 0.031 \left ( \frac{k_{\textup{B}}T}{\textup{keV}} \right )^{-1}\left ( \frac{M_{\textup{BH}}}{10^9~\textup{M}_{\odot}} \right )  
\end{equation}
where T is the gas temperature, and $M_{\textup{BH}}$ is the mass of the black hole. If we neglect angular momentum and assume an adiabatic index of $\gamma=5/3$, the rate at which the SMBH should accrete the surrounding hot gas at the Bondi radius may be found using: 
\begin{equation}
\frac{\dot{M}_{\textup{B}}}{\textup{M}_\odot~\textup{yr}^{-1}}= 0.012\left ( \frac{k_{\textup{B}}T}{\textup{keV}} \right )^{-3/2}\left ( \frac{n_e}{\textup{cm}^{-3}} \right )\left ( \frac{M_{\textup{BH}}}{10^9~\textup{M}_{\odot}} \right )^2
\end{equation}
where $n_e$ is the gas density.

Through the use of orbit superposition models, \cite{Thomas2016} determined the black hole mass for NGC 1600 to be $M_{\textup{BH}}=1.7\pm0.15\times10^{10}M_{\odot}$. The central temperature found from the single temperature profile is $1.20^{+0.15}_{-0.13}$~keV. These values give a Bondi radius of R$_{\textup{B}}= 0.38 - 0.54~\textup{kpc}~(1.2 - 1.7~\textup{arcsec})$. The temperature and density values for this range of Bondi radius gives a Bondi accretion rate of $\dot{M}_{\textup{B}} = 0.1 - 0.2~ \textup{M}_{\odot}~\textup{yr}^{-1}$. Furthermore, the Bondi accretion power, given by:
\begin{equation}
P_{\textup{B}}=\mu\dot{M}_{\textup{B}}c^2
\end{equation}
where $\mu$ is the efficiency, may be found assuming the conventional $10\%$ efficiency. Therefore, the Bondi accretion power is $P_{\textup{B}}=0.5-1\times10^{45}~\textup{ergs~s}^{-1}$. 

There appear to be a pair of cavities to the north and south of NGC 1600 (Left side of Fig.~\ref{fig:subimage}), each with a radius of around 1kpc. We therefore can calculate the power needed to create these cavities. The energy needed to create the bubble is given by $E=4PV$ (see e.g. \citealt{Birzan2004,Walker2014}), where P is the pressure of the surrounding ICM and V is the volume of the bubble. The power needed to inflate such a bubble is given by
\begin{equation}
    P = \frac{E}{t_{age}}
\end{equation}
where $t_{age}$ is the age of the cavity. Following for example \cite{Birzan2004} and \cite{Vantyghem2014}, we take $t_{age}$ to be the average of the sound crossing time, the buoyancy rise time, and the refill time. We calculate $t_{age}\approx10~Myrs$ and assume a cavity with $r\approx1.0$kpc, which gives a cavity power of $P=7.4\times10^{42}$ergs/s. The Bondi accretion power is therefore significantly higher than the power needed to inflate the cavities (as is also found for M87 in R15).
\section{Discussion}

\begin{figure*}
	\includegraphics[width=0.49\textwidth]{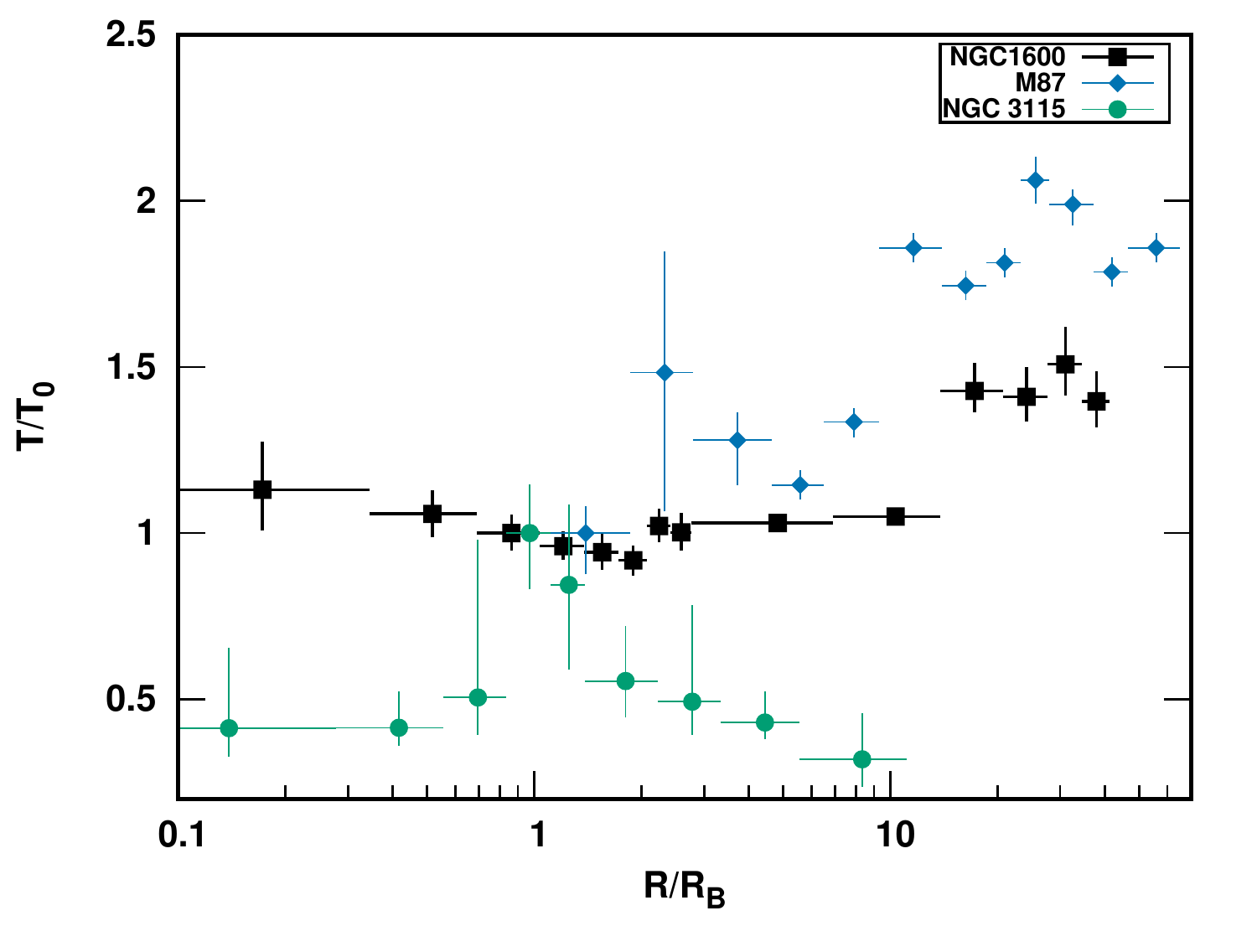}
	\includegraphics[width=0.49\textwidth]{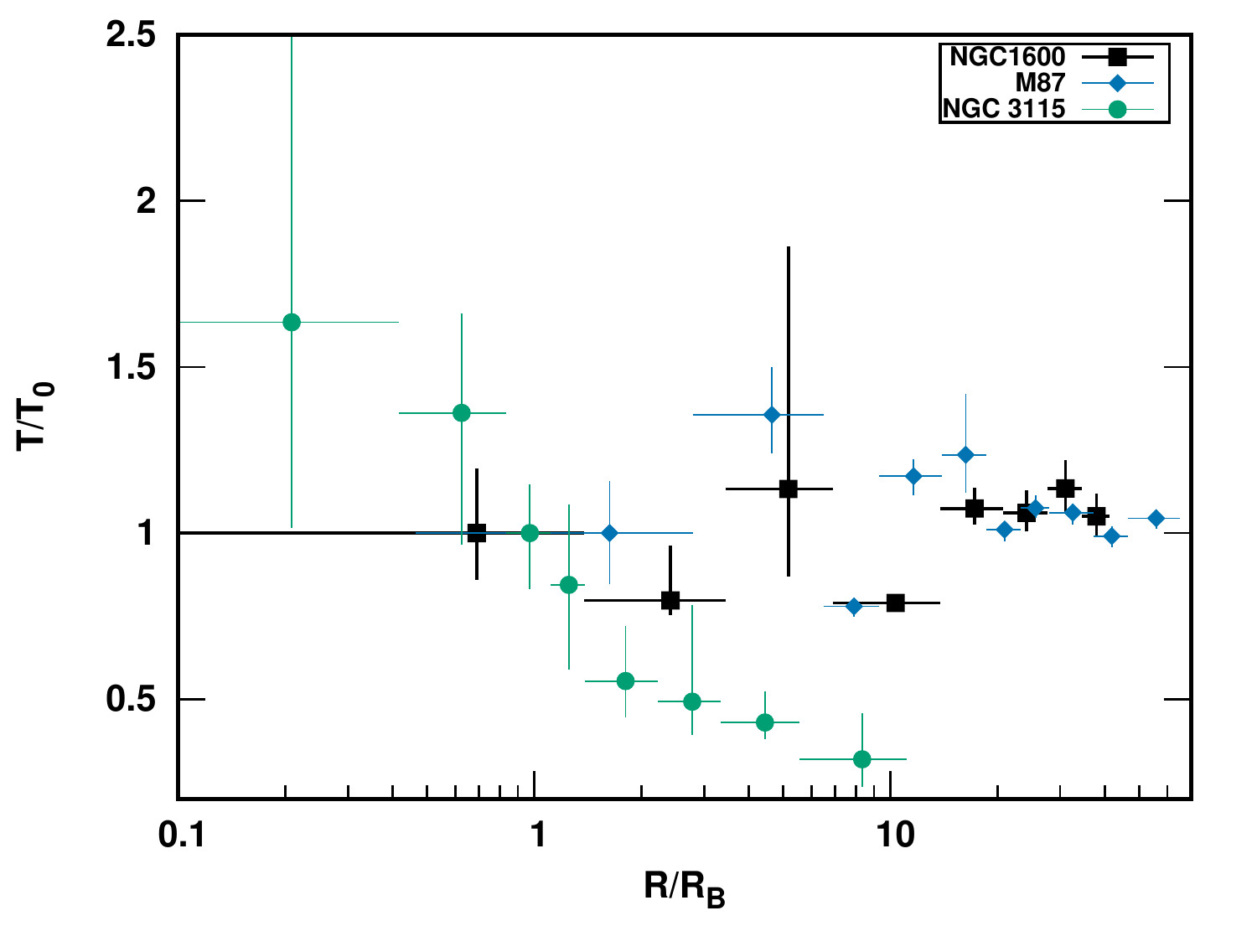}
	\caption{Left: Comparing the single-temperature profile of NGC 1600 to that of M87 (blue diamonds) and NGC 3115 (green circles). Here, R$_\textup{B}$ is the Bondi radius corresponding to each system and T$_{0}$ is the temperature at the Bondi radius. Right: Same as the left panel but using the temperature of the higher temperature (volume filling) component for the central regions where the gas has multiple temperature components. Only in NGC3115 does the temperature increase significantly as one moves towards the centre.}
	\label{fig:temp_density_compare}
\end{figure*}

\begin{figure}
	\includegraphics[width=0.49\textwidth]{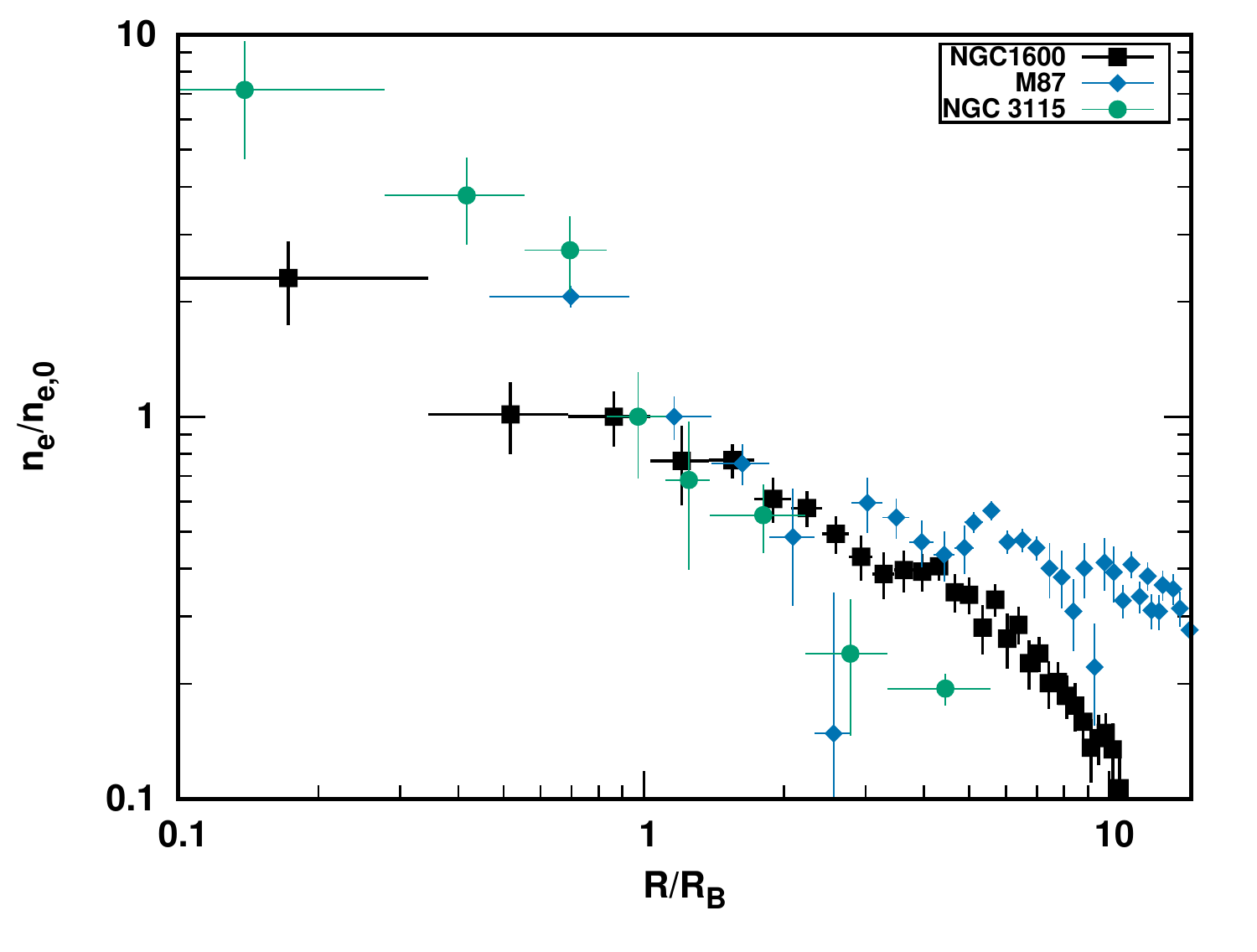}
	\caption{Comparing the deprojected density profile from the single temperature fits of NGC 1600 (black squares) to that of M87 (blue diamonds) and NGC 3115 (green circles). Here, n$_{e,0}$ is the electron density measured at the Bondi radius.}
	\label{fig:densityonly_compare}
\end{figure}

NGC 1600 is one of a small number of systems where the Bondi radius can be resolved using \textit{Chandra} observations, due in part to its unusually massive, central black hole. As \cite{Thomas2016} pointed out, NGC 1600 is an outlier in that it is not part of a rich galaxy cluster yet contains a black hole with mass on the order of $10^{10}\textup{M}_\odot$. The progenitors of such massive black holes are luminous quasars, and it has been shown at higher redshifts that the most luminous quasars are not biased when it comes to environment (\citealt{Trainor12,Fanidakis13}). Therefore, it is possible that NGC 1600 is a descendent of a luminous quasar outside a rich environment. We have calculated the temperature and density profile within the Bondi radius along with other important measurements, such as entropy and cooling time. Using these observations, a number of important characteristics of NGC 1600 can be discussed.

Both the single and two temperature model profiles reveal that the central temperature increases slightly within the Bondi radius, however this increase (a powerlaw of $T \propto r^{-0.073}$ for the 1 temperature fits) is much smaller than the $T \propto r^{-1}$ behavior expected from classical Bondi accretion models. This would appear to indicate that the dynamics of the hot gas are not determined by the ultramassive black hole, and that therefore the the Bondi accretion rate may not be an accurate estimate of the true accretion rate. A similar lack of a significant increase in temperature has also been found for M87 in \citet{Russell2015}. It is of course possible that there could be a temperature increase on scales smaller than the 0.1kpc scales that we can resolve with the current Chandra observations.

This contrasts with NGC 3115, also probed by \textit{Chandra}, where W14 find that there is an increase in temperature within the Bondi radius for the hotter gas component of the two temperature model ($T \propto r^{-0.44}$). Similar temperature increases have also been found in observations made of several elliptical galaxies (\citealt{Machacek06,Humphrey08,Pellegrini12}).

In Fig. \ref{fig:temp_density_compare} (left panel) we compare the temperature profiles found from the single temperature fits to NGC 1600 with the equivalent profiles for M87 and NGC 3115. To aid comparison (since the Bondi radii are different for all of these systems), we have scaled the radial axis by the Bondi radius and the temperature/density profiles by their values at the Bondi radius. In the right hand panel of Fig. \ref{fig:temp_density_compare} we compare the temperature profiles, but this time we use for the central multiphase regions the temperature of the high temperature component, which is interpreted by W14 and R15 as the volume filling component. As the right panel of Figure~\ref{fig:temp_density_compare} shows, the shape of the temperature profile of NGC 1600 more closely follows that of M87, and only in NGC 3115 is there a steeply rising temperature gradient. 

We find the density profile follows a $\rho \propto r^{-[0.61\pm0.13]}$ relation within the Bondi radius, which is very similar to the behavior found in NGC 3115, ($\rho \propto r^{-[0.62^{+0.26}_{-0.38}]}$), and slightly shallower than the case for M87 ($\rho \propto r^{-[1.0^{+0.2}_{-0.2}]}$). In a classical Bondi accretion flow \citep{Bondi52} the density is expected to vary more steeply for $r\ll \textup{R}_{\textup{B}}$ as $\rho \propto r^{-1.5}$.  
 Our observed, flatter central density profile is consistent with several analytical models and numerical simulations where the accretion rate decreases with radial distance (\citealt{SP01,HB02,YB10,Begelman12}). Simulations by \cite{Yuan12} predict a density profile of $\rho \propto r^{-(0.65-0.85)}$, which supports the ADIOS model where the inflow of mass at the Bondi radius is ejected from the black hole's sphere of influence, never reaching the event horizon (\citealt{BB99,BB04}). 

To allow a visual comparison of the shapes of the density profiles in the different systems, in Fig. \ref{fig:densityonly_compare} we compare the deprojected density profiles found from the single temperature fits to NGC 1600 with the equivalent profiles for M87 and NGC 3115, again scaling the radial axis by the Bondi radius, and the density by its value at the Bondi radius.

\subsection{Multiphase gas}

\begin{figure}
	\includegraphics[width=\linewidth]{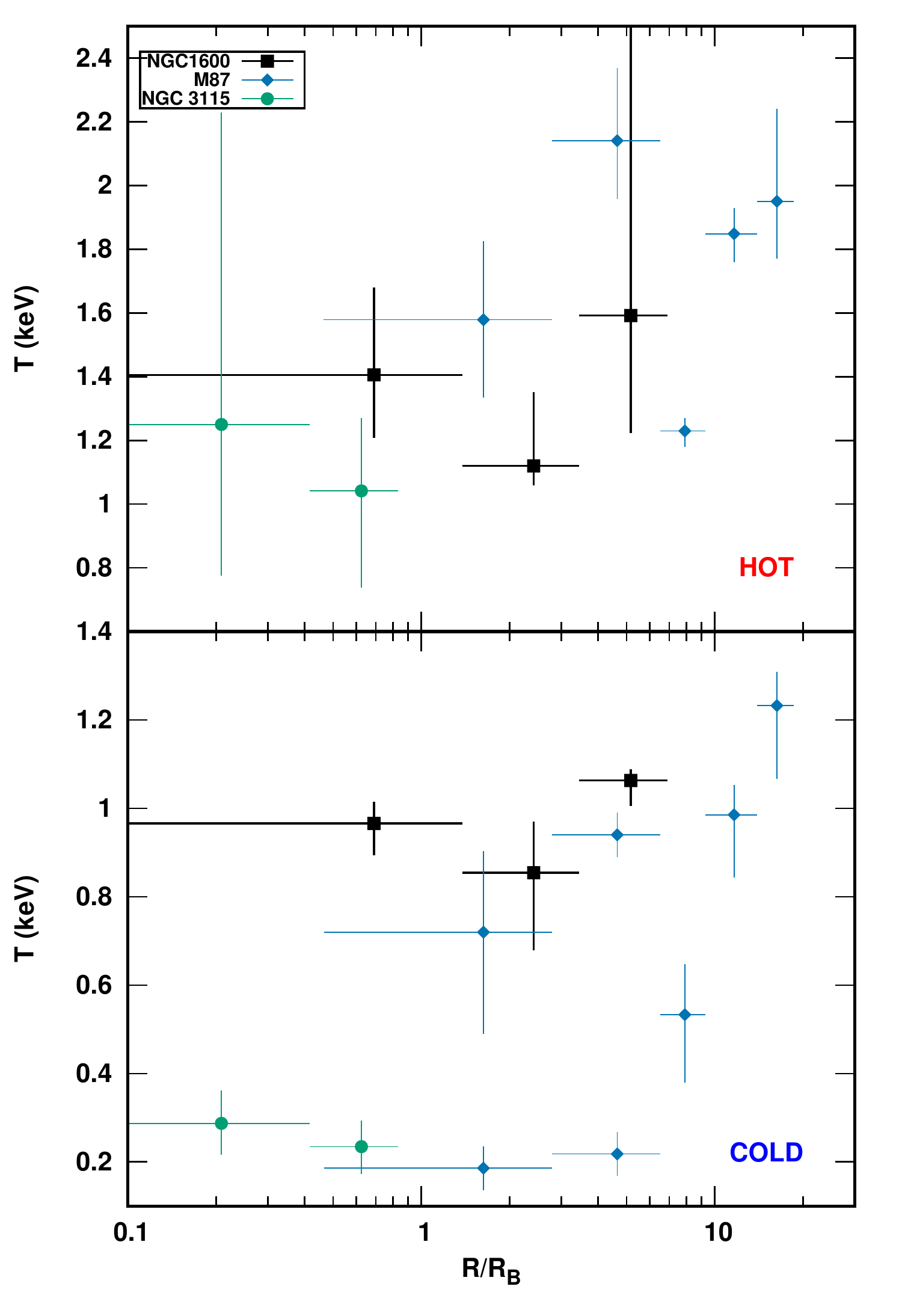}
	\caption{Top: Comparison of the hot temperature component from the multi-temperature model among NGC 1600, M87, and NGC 3115 with radial distance scaled by the Bondi radius. Bottom: Same as top but for the cold temperature components. In M87 there are 2 cold components, one at around 0.8keV and one at around 0.2keV.}
	\label{fig:2tempcomparison}
\end{figure}

Like W14 and \citet{Russell2015}, a two-temperature model was found to be statistically significant when fitting the spectra. With further deep observations of M87, \citet{Russell2018} also found an additional third very low temperature component (0.19keV) that could not be detected in the shallower observations studied in R15. 
In Figure~\ref{fig:2tempcomparison} we compare the temperatures of the hot and cold components found for NGC 1600 with those found for NGC 3115 and M87, scaling the radial axis by the Bondi radius. For NGC 3115, the gas only becomes multi-phase within the Bondi radius, whereas for NGC 1600 and M87 it is multiphase out to much larger radius, out to around 10-20 times the Bondi radius. For the high temperature component (top panel of Fig. \ref{fig:2tempcomparison}) all three are in relative agreement, with the innermost temperatures around the Bondi radius for all 3 systems in the range of 1-1.6keV.

The lower temperature component we find for NGC 1600 (Fig. \ref{fig:2tempcomparison}, bottom panel) is similar in temperature to the intermediate temperature component found in R18 for M87, in the range 0.6-1.1keV. The very low temperature component found by R18 for M87 at around 0.19 keV is not found in the shallower observations studied in R15, and we do not detect such a low temperature component for NGC 1600. As with M87, it is possible that with much deeper observations of NGC 1600 it will be possible to identify a low, 0.2keV component.

Systems with low cooling time and central entropy generally have H$\alpha$ emission, another indication of multiphase gas (\citealt{Cavagnolo08,Rafferty08,Voit15}). For NGC 1600, \citet{TdS91}, have reported an H$\alpha$ detection that extends out to $\sim\!10\arcsec$ or, $\sim\!3$~kpc, in good agreement with the regions we have found to be best fit with a two temperature model. 

 The onset of thermal instability has been shown to present itself in systems where the central entropy reaches a threshold of $\leq\!30~\textup{keV}~\textup{cm}^2$ (\citealt{Rafferty08,Cavagnolo08,Voit08}). The entropy values for NGC 1600 fall below this value within of $\sim3$~kpc from the center, in agreement with where with find the gas to become multi-phase. Additionally, theoretical studies and observations have shown that systems where the ratio of the cooling time to free-fall time falls below $t_{\textup{cool}}/t_{\textup{ff}}\leq10$, thermal instabilities are present (\citealt{McCourt12,Sharma12,Gaspari12,Voit15,VoitDonahue15,Loubser16}). The lowest value for NGC 1600 is $t_{\textup{cool}}/t_{\textup{ff}}=\!13.48\pm0.77$, not reaching the required ratio, but \cite{Voit16} show that this value may vary between $4<t_{\textup{cool}}/t_{\textup{ff}}<\!20$. However,~\cite{McNamara16} argue that it is the cooling time that overwhelmingly governs this property, where $t_{cool}<\!5\times10^8~\textup{yr}$ is the criteria. Looking at Figure~\ref{fig:time}, the cooling time measured for NGC 1600 falls below this value at $\sim\!3$~kpc, coincident with where the entropy falls below its threshold.

\section{Conclusions}
Using new deep \textit{Chandra} observations in conjunction with archival \textit{Chandra} data of NGC 1600, we have determined the temperature and density profiles within the Bondi accretion radius, down to a radius of $\sim\!0.16$~kpc from the central ultramassive black hole. We detect two, statistically significant temperature components within 3 kpc. The temperature profile increases very mildly within the Bondi radius, and is consistent with being flat. This contrasts with the expected increase in temperature towards the centre one would expect from classical Bondi accretion, suggesting the dynamics of the gas are not being determined by the central black hole. It is however, possible that the temperature increases on scales smaller than those that we can probe. This is similar to what is found for M87 in R15.

Based upon the temperature and density profiles, we calculate a mass accretion rate of $0.1 - 0.2~ \textup{M}_{\odot}~\textup{yr}^{-1}$ at the Bondi radius of R$_{\textup{B}}= 0.38 - 0.54~\textup{kpc}~(1.2 - 1.7~\textup{arcsec})$. Inside the Bondi radius, the density profile follows a power law of $\rho\propto~r^{-[0.61\pm0.13]}$, which is flatter than one would expect for classical Bondi accretion. This is consistent with observations of M87, NGC 3115, and SgrA$^*$ (\citealt{Russell2015,Wong2014, Wang2013}) and the numerical simulations of \cite{Yuan12} where $\rho\propto~r^{-(0.65-0.85)}$. Such an agreement favors models such as the ADIOS model where the mass inflow is ejected from the black hole's sphere of influence before reaching the event horizon.

Both the calculated entropy and cooling time profile support the multi-temperature model. The entropy drops below a critical value of $30~\textup{keV}~\textup{cm}^2$ within 3 kpc, which has been shown to be a common characteristic in systems with thermal instabilities. The cooling time reaches a threshold value for multiphase systems of $t_{cool}<\!5\times10^8~\textup{yr}$ at the same radius. Additionally, there is H$\alpha$ detection coincident with this low entropy, rapidly cooling gas, supporting our findings.

In our studies of the thermodynamic profiles within the Bondi radius, we have been limited to exploring azimuthally averaged values, similar to the initial study of M87 by \citet{Russell2015}. Future deeper observations of NGC1600 are necessary to explore the asymmetry of the density profile within the Bondi radius. 

Our results highlight the need for a next generation X-ray observatory with both high effective area and sub-arcsecond spatial resolution, such as the \emph{Lynx} \citep{Lynx} and AXIS \citep{AXIS} concepts, which will allow detailed, azimuthally resolved studies of the accretion flows within the Bondi radius.
\section*{Acknowledgements}
We thank the referee for their helpful report. JR and SAW acknowledge support from Chandra grant GO9-20073X. This work is based on observations obtained
with the Chandra observatory, a NASA mission.

\section*{Data Availability}
The Chandra Data Archive stores the data used in this paper. The \textit{Chandra} data were processed using the \textit{Chandra} Interactive Analysis of Observations (CIAO) software. The software packages \textsc{heasoft} and \textsc{xspec} were used, and these can be downloaded from the High Energy Astrophysics Science Archive Research Centre (HEASARC) software web page.

\bibliography{reflib}
\end{document}